\documentclass[11pt,twoside]{article}

\usepackage{asp2006}
\usepackage{epsfig}
\usepackage{lscape}

\begin{document}
\title{Recovering isolated galaxies from large scale surveys:
problems and strategies}
\author{Juan de Dios Santander-Vela,\altaffilmark{1,2,5}
Simon Verley,\altaffilmark{3,4,5} and
Lourdes Verdes-Montenegro\altaffilmark{1,5}}
\altaffiltext{1}{\emph{Instituto de Astrof\'isica de Andaluc\'ia-CSIC, Apdo. Correos 3004, E-18080 Granada, Spain}}
\altaffiltext{2}{\emph{European Southern Observatory, Karl-Schwarzschild-Str. 2, D-85748 Garching-bei-M\"unchen, Germany}}
\altaffiltext{3}{\emph{University of Granada, Dept. F\'isica Te\'orica y del Cosmos, Edificio Mecenas,  
Campus Universitario Fuentenueva, E-18071 Granada, Spain}}
\altaffiltext{4}{\emph{LERMA-OBSPM, 61 Av. de l'Observatoire, F-75014 Paris, FRANCE}}
\altaffiltext{5}{\emph{AMIGA Collaboration \texttt{http://amiga.iaa.es/}}}

\begin{abstract}
	The large survey programs being performed nowadays, being
	the SDSS their flagship, provide us with morphological
	parameters which allow for extraction of large galaxy samples.
	We will analyze the methodology for obtaining an AMIGA-like
	catalogue of isolated galaxies from the SDSS~DR5 photometric
	catalogue of galaxy objects, together with the roadblocks found
	in the process, and suggested workarounds.
\end{abstract}

%\section{}   %%% Top level section head
%\subsection{}   %%% Second level section head
%\subsubsection{}   %%% Lowest level section head (unnumbered)
%\section*{}    %%% Unnumbered top level section head
%\subsection*{}   %%% Unnumbered second level section head
\section{Introduction} % (fold)
\label{sec:introduction}
	The Catalogue of Isolated Galaxies (CIG) by
	\citet{1973AISAO...8....3K}, basis for the AMIGA sample
	\citep{2005A&A...436..443V}, was compiled using a visual
	search on photographic plates. In the era of the Virtual
	Observatory and large surveys, however, the automated
	compilation of isolated galaxy samples not only becomes
	feasible, but potentially less biased and error prone.
	
	 There are previous examples of automatic extraction of isolated
	galaxy samples in the literature, some of them mimicking
	Karachentseva's criterion, as is the case of
	\citet{2005AJ....129.2062A}. The search of compact groups uses
	similar constraints, and a recent example of automated
	extraction of them in galaxy catalogues is shown by
	\citet{2009MNRAS.395..255M}.
	
	 With automatically extracted samples, combined with the
	extra parameters derived by the SDSS pipeline, it is possible
	to estimate the degree of isolation of the selected galaxies,
	using for instance the tidal forces or local density
	estimations characterised in
	\citet{2007A&A...470..505V,2007A&A...472..121V}. Additionally,
	we gain the ability to re-do the studies, and apply them to
	similar catalogues in the southern hemisphere, so that an
	all-sky sample of isolated galaxies can be finally assembled.
	In this proceeding we will focus on the automatic extraction
	of a CIG (or AMIGA) like sample of isolated galaxies.
	
% section introduction (end)

\section{Implementing Karachentseva's isolation criterion on the SDSS} % (fold)
\label{sec:implementing_Karachentseva_s_isolation_criterion}
	
	A galaxy $i$, with angular diameter $D_i$ is considered isolated,
	following Karachentseva's criterion, 
	if the projected distance between it and neighbor $j$, $P_{ij}$,
	verifies
	
	\begin{equation}
		\label{isolation}
		P_{ij} \ge 20 \times D_j
	\end{equation}

	\noindent for all $j$ galaxies with diameter $D_j$,
	also verifying
	
	\begin{equation}
		\label{sizeComparison}
		\frac{1}{4} D_i \le D_j \le 4 \times D_i
	\end{equation}
	
	\citet{2005AJ....129.2062A} reformulates
	eq.~\ref{sizeComparison}, changing it for a relative
	magnitude difference comparison:
	
	\begin{equation}
		\label{magnitudeComparison}
		\left|g_i - g_j\right| > 3.0 
	\end{equation}
	
	\noindent where $g_i$ is the $g$ magnitude of galaxy $i$;
	both approaches
	are equivalent assuming
	constant surface brightness profiles. We tried both
	approximations in our testing.
	
	 As the SDSS spectrometric catalogue is known to be complete only
	up to Petrosian magnitude $r$ brighter than
	17.7~\citep{2007ApJS..172..634A}, we will make use of
	the SDSS photometric catalogue of objects identified as galaxies.
	We targeted SDSS~DR5, but as the SDSS table
	schemata for DR7 have only added columns to specify clean
	objects, our procedure is valid also for SDSS~DR7.
	
	Using CasJobs,
	we selected objects from the \texttt{PhotoObj}
	table view with
	\texttt{type~=~3} (Galaxy), and $g$ magnitudes between 
	10 and 15. 
	For convenience during the testing, we applied a RA/Dec cut:
	
	\begin{eqnarray*}
		206.6  < &  \mathrm{RA}   < & 228.5\\
		 31.4  < &  \mathrm{Dec}  < &  35.5
	\end{eqnarray*}
	
	\noindent which gave us a test field containing around 98000
	objects, with 2 CIG galaxies (CIG~613 \& CIG~642) among them.
	
	After downloading the test field data, 
	a Python script was run to check galaxy isolation,
	marking galaxy $i$ as not isolated if any galaxy within
	80 times $D_i$ in the test frame (the maximum relevant distance,
	giving the size comparison) which verified
	eq.~\ref{sizeComparison}
	did not verify eq.~\ref{isolation}.
	A second script
	%did the same, but using
	used eqs.~\ref{magnitudeComparison} and
	\ref{isolation} instead.
	
	The two approaches resulted to be largely equivalent, but
	the results gave a lot more allegedly isolated galaxies than
	the two CIGs. If the factor of 20 in
	eq.~\ref{isolation} was increased, the results
	provided none of the two CIGs, and retained some objects
	in the center which did not appear to be isolated galaxies,
	plus objects along the frame border (an artifact of the reduced
	sample).
	
% section retrieving_isolated_galaxies_from_sdss (end)

\section{Problems found and strategies to avoid them} % (fold)
\label{sec:problems_found_and_strategies}
	
	In order to understand why the two CIGs in the test frame
	could not be recovered as isolated, we checked visually the
	frame and found several problems, mainly the misclassification
	of stars as galaxies, and item blending.
	
	 As we intend to recover a catalogue of
	truly isolated galaxies, without resorting to visual inspection
	but at the latest stage,
	we need to find a way to objectively distinguish stars which
	have been misclassified as galaxies by the original SDSS
	pipeline\footnote{The original SDSS photometric star-galaxy separation is
	done using the difference between modeled profiles (linear
	combination of the best-fitting De~Vaucouleurs and exponential
	profiles) and PSF model magnitudes.}.
	
	\begin{figure}[tbp]
		\centering
		\includegraphics[
			width=0.9\textwidth,
			clip,
			viewport= 0.30in 0.28in 6.34in 1.99in
		]
		{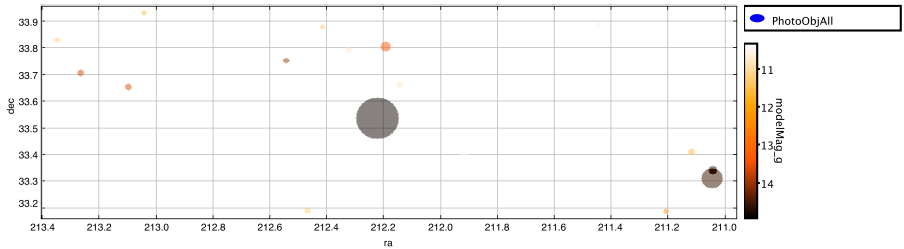}
		\caption{
			Object positions, radii, and magnitudes for in subset of
			the test field around CIG~613 (center object). Magnitude
			is shown in the color scale (darker is less bright),
			while radii are proportional to the Petrosian radius for
			each object.
		}
		\label{fig:SubField}
	\end{figure}
	
	Figure~\ref{fig:SubField} shows how CIG~613 (the central, largest
	dark circle) is bigger than the surrounding objects, but less
	bright than most companion objects, which exhibit a much higher
	surface brightness. This shows that there are parameters, or
	combinations of parameters, helpful to objectively improve
	SDSS' star-galaxy separation. In fact, such kind of star-galaxy
	separation had already been independently performed by
	\citet{2007MNRAS.375...68C}, disregarding SDSS \texttt{type}
	entirely, based also on additional information from the SDSS
	photometric catalogue itself. However, their star-galaxy
	separation algorithm includes photometric redshift estimation,
	which is not reliable in the the local universe, due to the
	strong influence of peculiar velocities.
	
	 Of the large number of columns available for
	\texttt{PhotoObj}\footnote{
	\texttt{http://cas.sdss.org/dr5/en/help/browser/description.asp?n=PhotoObj\&t=V}
	In particular,
	\texttt{probPSF} is the probability of the object being a star;
	\texttt{lnLStar} is the $ln$ of star fitting likelihood; \texttt{lnLExp} and \texttt{lnLDeV} are the $ln$
	of the likelihood for the object fitting an exponential disk or a De~Vaucouleurs profile.
	These parameters are available for each $ugriz$ color.},
	several provide information about the
	reliability of the SDSS pipeline's star/gal\-a\-xy separation, such as
	\texttt{probPSF}, \texttt{lnLStar}, plus other
	parameters such as \texttt{lnLExp} ,
	\texttt{lnLDeV}, and many others. However, many of them
	showed binary, non-continuous values, and where not useful for
	building a separation. The most promising
	single parameter, able by itself to perform an effective
	separation, is \texttt{lnLStar\_g}. The lowest absolute value of \texttt{lnLStar\_g} for
	a CIG galaxy was higher than 29100 (for CIG 299), with the rest
	of CIG's having $abs(\mathtt{lnLStar\_g})$ above 40000.
	
\begin{figure}[tbp]
		\centering
			\includegraphics[
				width=0.9\textwidth,
				clip,
				viewport= 0.30in 0.28in 6.34in 1.99in
			]{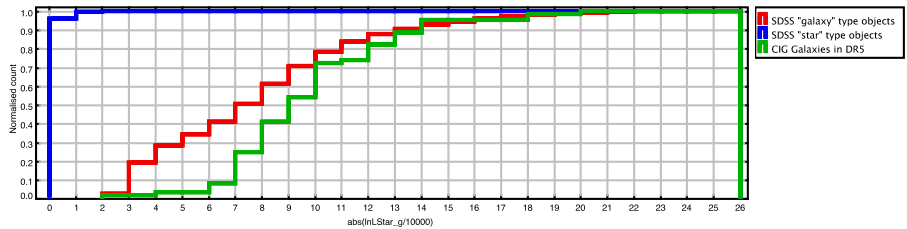}
		\caption{
			Normalized cumulative frequency histogram for the
			absolute value of \texttt{lnLStar\_g}, both for SDSS
			galaxies with $10 \le mag_g \le 15$ (red), CIG galaxies
			with SDSS~DR5 counterparts (green), and SDSS stars with the
			same magnitude cut (blue).
		}
		\label{fig:cumulHistoLNStar}
	\end{figure}

	This is shown in figure~\ref{fig:cumulHistoLNStar}, where we
	plot the normalized
	distribution of
	$abs(\mathtt{lnLStar\_g})$ values across the SDSS~DR5 galaxies of our
	test field
	(red line), and the CIG galaxies with SDSS~DR5 counterpart
	(green). For comparison, we have plotted also the histogram
	for SDSS star objects
	(\texttt{type~=~6}; in green), and we can see that the absolute
	value of \texttt{lnLStar\_g}
	is below 10000 for more than 95\% of star objects.

% section problems_found_and_strategies (end)

\section{Conclusions and Future Work} % (fold)
\label{sec:conclusions}
	
	We have seen that in order to use the SDSS Photometric catalogue
	for retrieving galaxy properties, and estimate their isolation
	degree, the basic star-galaxy separation provided by the SDSS
	pipeline is not enough, and more work is needed in order to be
	able to obtain meaningful results regarding the isolation of
	galaxies.
	
	 The analysis of available parameters, however, has shown to be
	promising in achieving a more accurate star-galaxy separation
	which, combined with data mining and clustering tools, may
	finally allow a reuse of SDSS photometric data for this kind of
	studies.
	
	 In the future, we will continue pursuing an automated isolated
	galaxy sample builder system, using data mining techniques in
	order to find out the parameters with the best star-galaxy
	separation properties, and be able to extend the selection of
	isolated galaxies to similar catalogues in the future.
	%We plan
	%to make both our catalogue, and the final SDSS exploitation
	%tools, available through the Virtual Observatory.
	
% section conclusions (end)

\acknowledgements The authors wish to acknowledge the support
from DGI Grant AYA2008-06181-C02, and Junta de Andalucía
(Spain) grants TIC-114 and P08-FQM-4205-PEX. In addition, SV is
supported through the Spanish MICINN's Juan de la Cierva
post-doctoral program.

\end{document}